\documentclass[preprint2]{aastex}

\begin{document}

\title{Measurement of the Parallax of PSR B0950+08 Using the VLBA}

\author{Walter F. Brisken\altaffilmark{1},
	John M. Benson\altaffilmark{2}, 
	A. J. Beasley\altaffilmark{3}, 
	Edward B. Fomalont\altaffilmark{3}, 
	W. M. Goss\altaffilmark{2},
	S. E. Thorsett\altaffilmark{4}}

\altaffiltext{1}{Dept. of Physics, Princeton University, Princeton, NJ 08544,
email: walterfb@nacho.princeton.edu }
\altaffiltext{2}{NRAO, P.O. Box O, Socorro, NM 87801}
\altaffiltext{3}{NRAO, Edgemont Road, Charlottesville, VA 22903-2475}
\altaffiltext{4}{Dept. of Astr. and Astroph., University of California, 
	Santa Cruz, CA 95064}

\begin{abstract}
A new technique has been developed to remove the ionosphere's distorting
effects from low frequency VLBI data.  By fitting dispersive and 
non-dispersive components to the phases of multi-frequency data, the
ionosphere can be effectively removed from the data without the use of 
{\em a priori} calibration information.  This technique, along with the
new gating capability of the VLBA correlator, was used to perform accurate
astrometry on pulsar B0950+08, resulting in a much improved measurement of
this pulsar's proper motion ($\mu_{\alpha} = -1.6 \pm 0.4$ mas/yr, $\mu_{\delta}
= 29.5 \pm 0.5$ mas/yr) and parallax ($\pi = 3.6 \pm 0.3$ mas).  This puts the
pulsar at a distance of $280 \pm 25$ parsecs, about twice as far as previous 
estimates, but in good agreement with models of the electron density in
the local bubble.
\end{abstract}


\section{Introduction}

Accurate radio pulsar distances are important for population and birthrate
modeling and, combined with measurements of dispersion and Faraday
rotation, for studies of the ionized interstellar medium and Galactic magnetic
fields.  Parallax measurements can firmly fix the bottom rungs of the pulsar
distance ladder, but parallaxes of even the closest pulsars are only a few
milliarcseconds in amplitude, and such measurements are available for only 
nine pulsars (see
Toscano et al.\ 1999 for a recent list).\footnote{An optical parallax is
also available for the pulsar B0633+17 (Geminga), which has no known radio
counterpart.}

Four of these parallax measurements were made using timing studies of
millisecond pulsars with particularly stable spin periods.  
The others were made with various VLBI techniques.  The use of calibration
sources within the primary antenna beam shows promise, as demonstrated by 
\citet{fom99}.  Observing at L~band with 25 meter telescopes provides a primary beam size of about 30 minutes of arc.  Only a small fraction of pulsars have
calibrators within this small distance.
The number of accessible target pulsars is greatly increased by the
use of calibrators at greater offset angles, up to several degrees,
but differential ionospheric
phase delays make high precision comparison measurements very difficult.

Most pulsars are weak radio sources at high frequencies,
where array resolution is high and ionospheric effects are minimized.
Observations are often made at L~band (1.4 to 1.7 GHz)
as a compromise, but even at L~band the ionosphere remains a significant
obstacle.  We have developed a new calibration technique to remove the effects
of the ionosphere from the data and used it to measure a revised parallax of
PSR B0950+08.

\section{Observations}

At L~band, the VLBA provides simultaneous access to a large spanned 
bandwidth and is sensitive enough to detect a pulsar within
the changing ionosphere's timescale of about one minute.  Recently,
gating the VLBA correlator became possible.  This allows the correlator to
be turned off during the off pulse phase of the pulsar, increasing the
signal to noise of the pulsar by a factor of 3 to 6, depending on the pulse
duty cycle.  For PSR B0950+08 the improvement was a factor of 3.6.

We observed the pulsar simultaneously in eight 8 MHz bands ranging from
1410 MHz to 1730 MHz.  Each band was split into thirty two 250 kHz channels.
The VLBA antennas nod back and forth between the pulsar and the calibrator with
cycle times of about five minutes.  Seven epochs from 1994 to 1999 were observed
near parallax extrema.  The first three epochs were ungated and used 1004+141 
as a calibrator
source.  Its distance of seven degrees from the pulsar made phase
referencing difficult.  These data were useful for developing techniques and
understanding limitations but 
were not of high enough quality for accurate astrometry.  
The more recent epochs used J0946+1017, less than three
degrees distant from PSR B0950+08, as a calibrator.

Most astrometry at low frequency is done at night since the ionosphere phase
fluctuations are two
or three times stronger during the day.  Since parallax measurements are most
effective
when the target and the sun are 90 degrees apart in the sky, these observations
need to be done at sunrise and sunset---the times when the ionosphere 
changes most rapidly.

\section{Data Reduction}

\subsection{Phase-referencing}

This is a VLBI phase-referenced \citep{tay99} experiment with an additional
calibration step.  
A strong source, J0946+1017, was used as the reference source.  Using this
object's known position, the fringe rates and delays were measured and 
interpolated over the pulsar scans.  The calibrator is self-calibrated so 
that its structure can be modeled.  The self-calibration phases are 
interpolated over the pulsar.
This simple phase referencing determines the pulsar's position to 
less than 30 mas.  The phase center is shifted to the pulsar's position
and the 32 spectral channels within each 8 MHz band
are averaged together.  The large number of channels 
prevents bandwidth smearing from decorrelating the
long baseline visibilities.

\subsection{The Ionosphere}

\begin{figure}[h]
\plotone{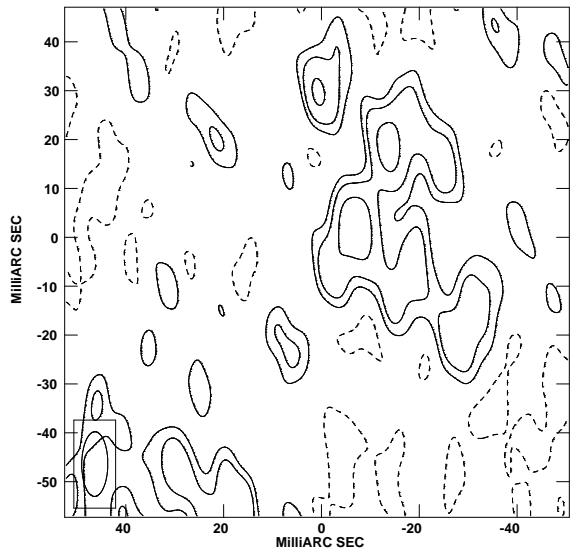}
\caption{Image of pulsar before removing the ionosphere.  All of the flux is
from the pulsar, scattered by the ionosphere.  The beam is 4 by 11 mas.  The 
lowest contour level is at 5 mJy and a factor of 2 separates contours.  All 
data from all telescopes were used to produce this image.}
\label{fig:withion}
\end{figure}

The ionosphere, like the free electrons in the ISM, introduces a frequency
dependent delay to the radio waves.  Unlike for the ISM, this
delay can be drastically different above each antenna since the
ionosphere has structure on all scales, ranging from meter-sized variations to
the bulge caused by the sun's radiation which stretches a third
of the way around Earth.  The ionosphere is also different along different
lines of sight from the same antenna.  Since the lines of sight from each
telescope to the pulsar and calibrator puncture the ionosphere at different
points, phase-referencing only calibrates away the
bulk ionosphere over each antenna, leaving spatial gradients in the 
ionosphere
uncorrected.  These gradients can change significantly on time scales 
shorter than one minute and are large enough to move the pulsar by as much as 
10 mas per degree of source separation.  
A phase referenced image with no ionospheric correction is shown in
figure \ref{fig:withion}.

At a given time on a given baseline, the phase-referenced
visibility phases have this structure:
\begin{eqnarray*}
\phi(\nu) = \underbrace{\left[
        \frac{1}{c}\left(U \; \ell + V \; m\right) + \delta A
\right] \nu}_{\mbox{non-dispersive}} +
\underbrace{\delta B\frac{1}{\nu}}_{\mbox{dispersive}}
\end{eqnarray*}

\noindent
$U$ and $V$ are the projected baseline lengths, $\ell$ and $m$ are the
pulsar's vector distance from the phase center (the quantity that this
experiment is trying to measure), and $\delta A \; \nu$ is the phase due to the
unmodeled wet troposphere and is very small, introducing less than 45 degrees
of phase.  These three terms are
non-dispersive and hence have the same frequency dependence.  $\delta A$ is measured in rad/GHz, and $\delta B$ is in
rad$\cdot$GHz.  $\delta B / \nu$
is the phase due to uncorrected ionosphere.  Since the remaining ionosphere
is the only dispersive term, it can be separated from the
non-dispersive sources of phase.

This separation, however, is complicated by phase ambiguity.  For a given set
of phase measurements, a whole family of solutions can be formed since
$n \cdot 2 \pi$ can be added to the original phases.  Because our bandwidth is
limited, any integral value of $n$ will generate a different solution with
an almost equally good fit.  It is fortunate that the 
value of $\delta A$ is always small enough that a
good guess of the pulsar's position constrains the ambiguity to a single
value of $n$.

\subsection{Determining the Pulsar's Position}

\begin{figure}[h]
\plotone{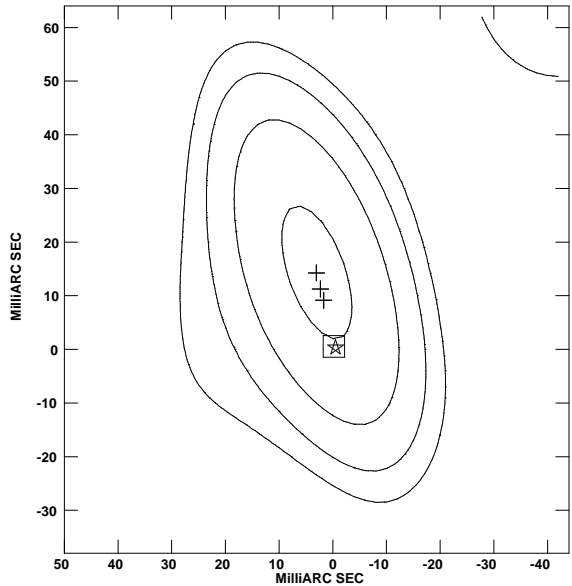}
\caption{30 minute snapshot made at 1.40 GHz using only the 5 Southwestern
antennas.  The beam size and shape match that of the second-highest
contour.  The three pluses designate measured location of the pulsar
at three different frequencies; from the top, these frequencies are: 1.40 GHz,
1.63 GHz, and 1.73 GHz.  The star indicates the inferred true location of the
pulsar by extrapolating the measured positions to infinite frequency.  The
center of the square is where the pulsar was determined to be using the
techniques shown in this paper.}
\label{fig:IF1}
\end{figure}

\begin{figure}[h]
\plotone{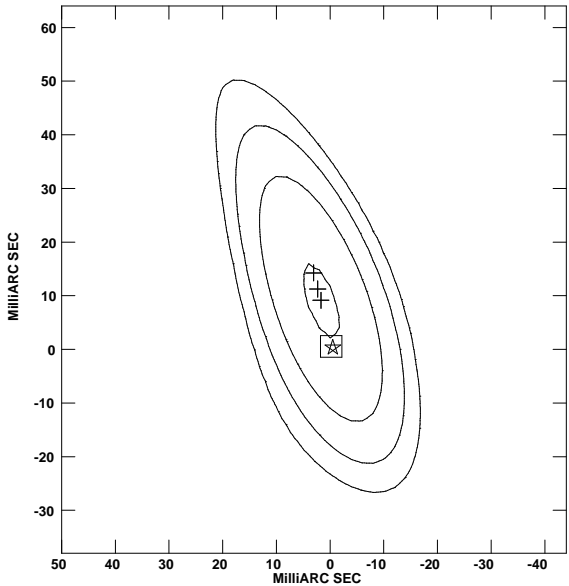}
\caption{Same as figure \ref{fig:IF1} but at 1.73 GHz.}
\label{fig:IF8}
\end{figure}

A bootstrapping approach is used to determine the pulsar's position.   
Using just the 5 Southwestern VLBA antennas ( Los Alamos NM, Pie Town NM,
Fort Davis TX, Kitt Peak AZ, and Owens Valley CA ) with a maximum
baseline of 1500 km, a separate image is made for each 8 MHz
band in the phase-referenced data.  Figures \ref{fig:IF1} and \ref{fig:IF8}
show two such images.  Plotting the measured positions as a function
of frequency reveals a trend -- the points line up in
frequency order.  If we assume a simplified model of the ionosphere as
a single wedge lying above all the antennas then we would expect 
the pulsar to appear at $\vec{P}(\nu) = \vec{P_{\infty}} + \vec{w} /\nu^2$. 
Fitting for $\vec{P_{\infty}}$, the extrapolated
infinite frequency position, yields an initial guess of the pulsar's
position ( the star on figures \ref{fig:IF1} and \ref{fig:IF8} ) with a 
precision of 3-10 mas.  The wedge model is certainly far from
correct over the whole Earth but works fairly well for the Southwest array.
Because the ionosphere changes so quickly,
a 15 to 30 minute snapshot usually works better than imaging all of the data.

\begin{figure}[h]
\plotone{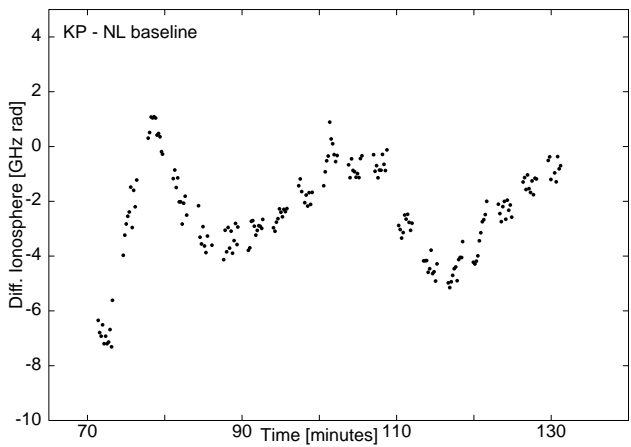}
\caption{Time variation of $\delta B$ on the Kitt Peak to North Liberty 
baseline.}
\label{fig:kpnl}
\end{figure}

\begin{figure}[h]
\plotone{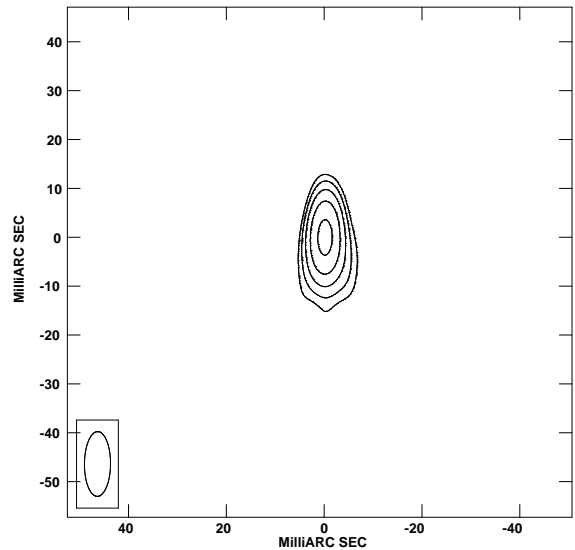}
\caption{Image of the pulsar after removing the ionosphere.  The same region
as figure \ref{fig:withion} is shown with the same contour levels. The highest
contour is at 64 mJy}
\label{fig:withoution}
\end{figure}

To get antenna-based phases as a function of frequency, the pulsar is
self-calibrated with a solution interval of between twelve and thirty seconds
and the antenna-based phase corrections are recorded.  The
self-calibration phases cannot be applied to the pulsar, as that would move
the pulsar to the phase center.  For each of the Southwestern antennas, the
nondispersive and dispersive components of the self-calibration solutions are 
fit, holding the geometric $\ell$ and $m$ terms fixed.  This is reasonable
since we know the pulsar's location to much less than 30 mas, the fringe spacing
of the longest Southwestern baseline.
The other antennas are initially ignored as the phase contribution due to 
geometric delay is significant on the longer baselines.
A new calibration table is made containing only the dispersive $\delta B/\nu$
corrections that were fit to the data. 
Imaging the data with these corrections produces a new position accurate to
less than a milliarcsecond.
With the improved pulsar position, the fitting for $\delta B$ and imaging are 
repeated with all antennas.  The final 
calibration table contains phase corrections for each antenna tabulated 
at time intervals of a minute or less.  Figure \ref{fig:kpnl} shows
the $\delta B$ values for the Kitt Peak-to-North Liberty baseline.  
This final image has much less scattered power and all frequencies appear at
the same location (figure \ref{fig:withoution}).

In summary, although the self-calibration process is used to derive 
antenna-based phases over short periods of time, only the dispersive part of
these phase terms is removed, removing the ionospheric distortions.
The non-dispersive part of the phases which contain information about
the source position (as well as small atmospheric terms) remain in the data.

\section{Application of Ionospheric Removal to PSR B0950+08}

PSR B0950+08 has a published parallax of $7.9 \pm 0.8$ mas \citep{gwinn}.
This measurement was made with a three element VLBI array consisting of Arecibo,
the 140 ft. antenna at Green Bank and the 40 m antenna at Owens Valley.  
A 3-element interferometer
produces a beam pattern with severe lobe ambiguity.  To break the degeneracy
of the solution, the proper motion was used to select
a self-consistent set of lobes.  Unfortunately, the published values of
$\mu_{\alpha} = 
15 \pm 8$ mas/yr and $\mu_{\delta} = 31 \pm 4$ mas/yr \citep{las82} were 
not sufficiently accurate, causing an incorrect choice of lobes. 

\begin{figure}[h]
\plotone{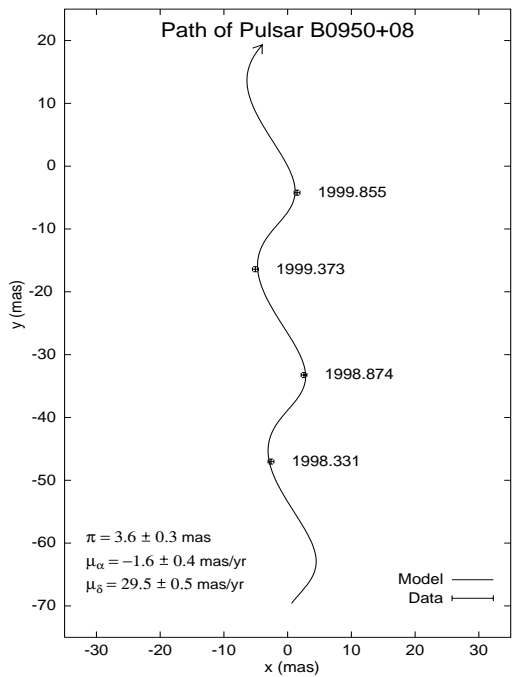}
\caption{Location of pulsar at four epochs and model fit to the points.}
\label{fig:motion}
\end{figure}

The four gated epochs of data were calibrated as described and fit with a
model with five components: position of pulsar (two components),
proper motion of pulsar (two) and parallax (one).  The fit is shown in figure
\ref{fig:motion}, and the proper motion and parallax are:
\begin{eqnarray*}
\mu_{\alpha} & = & -1.6 \pm 0.4\;\mathrm{mas/yr} \\
\mu_{\delta} & = & 29.5 \pm 0.5\;\mathrm{mas/yr} \\
\pi & = & 3.6 \pm 0.3\;\mathrm{mas}
\end{eqnarray*}
\noindent
The first three epochs of data suggest a parallax of $4 \pm 1$ mas suggesting
some stability of correction even with a calibrator separation of 7 degrees.
The nutation model in the correlator may leave an annual signature with
an amplitude of about 40 $\mu$as per degree of separation \citep{ma98}, 
about 0.1 mas in this experiment.  
Systematic errors such as these are not included in the error bars but are 
expected to be less than 0.3 mas.  

\section{Implications}

There has been a long-standing problem of reconciling the electron
density derived from the distance to PSR B0950+08 with that based on X-ray 
data \citep{tos99}.  
This measurement puts the pulsar more than twice as far away,
changing the local electron density from $n_e = 0.023 \mbox{cm}^{-3}$ to
$n_e = 0.01 \mbox{cm}^{-3}$,
a value much closer to the estimated electron density inside our local bubble,
$n_e \approx 0.005 \mbox{cm}^{-3}$.

\acknowledgments

The National Radio Astronomy Observatory is a facility of the National Science 
Foundation operated under cooperative agreement by Associated Universities, 
Inc.  This reasearch was funded in part by the National Science Foundation
including a graduate student fellowship.  Additional funding was provided by
Princeton University.  S.E.T. is a Sloan Research Fellow.

\end{document}